\documentclass[aps,preprint,showpacs,amsmath,amssymb]{revtex4}

\usepackage{graphicx}
\usepackage{dcolumn}
\usepackage{bm}
\usepackage{color}
\usepackage{graphicx}

\usepackage{slashed}
\usepackage{amssymb}
\usepackage{array}

\begin{document}

\title{Isospin Effect in Three-Body Kaonic Clusters}
\author{I. Filikhin}
\email{ifilikhin@nccu.edu} 
\affiliation{North Carolina Central University, Durham, NC 27707, USA}
\author{B. Vlahovic}
\affiliation{North Carolina Central University, Durham, NC 27707, USA}

\date{\today}

\begin{abstract}
The kaonic clusters $K^{-}K^{-}p$ and $ppK^{-}$ are  described  
 based on the configuration space Faddeev equations for $AAB$ system. The  $AB$ interaction is given by
 isospin-dependent potentials. For this isospin model, we 
show that  the relation  $\left\vert E_{3}(V_{AA}=0)\right\vert~<~2\left\vert
E_{2}\right\vert$ is satisfied when $E_{2}$ is the binding energy of the $AB$
subsystem and  $E_{3}(V_{AA}=0)$ is the three-body binding energy  when interaction between identical particles is omitted,  $V_{AA}=0$.
For the $NN{\bar K}$ system,  taking into account weak attraction of  $NN$ interaction  the relation leads to the evaluation $|E_3|\le 2|E_2|$. The ''isospinless model'' for the kaonic clusters based on the isospin averaged $N{\bar K}$ potential demonstrates the opposite relation  $\left\vert E_{3}(V_{AA}=0)\right\vert~>~2\left\vert E_{2}\right\vert$.
The isospin ''given charge formalism''  is presented for  $NN{\bar K}$ cluster. This formalism is related to  isospin model by unitary transformation of the isospin basis. An interpretation of the ''particle representation'' for $NN{\bar K}$ system is proposed.
\end{abstract}
\index{}\maketitle
 
%
\section{Introduction}
\label{intro}
The quasi-bound states in the kaonic clusters  $NN{\bar K}$ and ${\bar K}{\bar K}N$
are intensively debated during the last years. 
The main problem is
that  theoretical evaluations for the binding energy are in significant disagreement  with the values derived from existing experimental data \cite{G2016}.
The properties of the kaonic clusters are defined by  ${\bar K}N$ interaction, having significant difference for the isospin singlet and triplet channels. 
The isospin singlet component of the ${\bar K}N$ potential generates a quasi-bound state corresponding to the $\Lambda(1405)$ resonance  below the $K^-p$ threshold \cite{HAY2017}.
The resonance has the double state nature due to  the ${\bar K}N$ quasi-bound state  and $\pi\Sigma$ resonance \cite{D1960,YA07,J10}. 
One can point out two models for the ${\bar K}N$ quasi-bound state  to be used within three-body calculations. 
The AY ${\bar K}N$  potential effectively taking into account the  $\pi\Sigma$  coupling  has been proposed in Ref. \cite{YA07}.
The effective 
${\bar K}N$ interactions have the strong attraction in the singlet $I=0$  channel 
and the weak attraction in the triplet $I=1$ channel. 
The $ppK^-$ binding energy  obtained within this  model is $|E_{NN{\bar K}}|$=48~MeV. Two-body threshold
is close to the bound state energy of $\Lambda(1405)$ as $K^-p$ bound pair (about 30~MeV).
Similar results have been obtained within more complex models \cite{D17,DIM2015,S15,RS14,MKE16,SGM07}.
This value is much smaller than the experimentally motivated value of about 100~MeV for the $ppK^-$ ''deeply bound state''\cite{A05,Y10,I15}. 
Alternatively, the chiral model for the  potential has been proposed (see \cite{HW,DHW08,DHW09}) for the ${\bar K}N$ interaction. The model
reduces the singlet component of ${\bar K}N$ potential due to the strong coupling  ${\bar K}N$  and $\pi\Sigma$ channels
which give contribution $3/1$ in three-body amplitudes, respectively.
The value about 20~MeV for  $|E_{NN{\bar K}}|$  was obtained with the two-body threshold about 12~MeV. This model
is accompanied by the energy dependence  of  ${\bar K}N$ interaction and includes the $p$- wave component of ${\bar K}N$ potential into consideration. The coupling between the channels  $NN{\bar K}(S_{NN}=0)$ and   $NN{\bar K}(S_{NN}=1)$ is also taken into account. Note that these factors affect three-body binding energy in 1-7 MeV and contribute with different sings. The effect of the energy dependence is discussed \cite{R17}.

Discussion about the experimental background  and theoretical interpretations can be found in Ref. \cite{GHM,G2016,Na2016,SOR18}.

In the presented work we consider  the lower bounds for the  ground state energy of   three-body kaonic clusters
appeared due to strong  isospin dependence of  ${\bar K}N$ potential when  ${\bar K}N$ pair is deeply bound and  $NN$ potential is  relatively weaker.
For the  $NN{\bar K}$ cluster, we show that  
the ''AY-like'' models \cite{AY02} cannot leads to ''$ppK^-$ deeply bound state'' which was assumed to be taking into account in the existing experimental data treatments. 

Our study is based on the Faddeev equations in configuration space \cite{FaddeevConfigurSpace}.  The Faddeev equations allow to
separate components of the total wave function corresponding to the different particle rearrangements
 and evaluate the contribution of each configuration.

We consider the $NN{\bar K}$ and ${\bar K}{\bar K}N$ clusters as  three-body $AAB$ systems that include two identical particles $AA$ and
study the relation between the ground state energies of the $AB$
subsystem, $E_{2}$, and  three-body system, $E_{3}(V_{AA}=0)$, when the interaction between the identical particles is omitted.
The kaonic clusters are the systems having  isospin dependent $AB$ interaction. The triplet and singlet components of $N{\bar K}$ interaction are essentially different.
We show, that for such systems, the  relation $ |E_3(V_{AA}=0)|<|2E_2| $ takes place.
Based on this relation, we can formulate more strong statement: $ |E_3|<|2E_2| $,  taking into account weak attractive (or weak repulsive)
$AA$ potentials.  
The last relation shows that $NN{\bar K}$ ''AY-like'' calculations for $|E_3|$ have to result in the values smaller than 60~MeV.

The relation between  $E_{2}$ and  $E_{3}(V_{AA}=0)$ has been previously considered  for bosonic-like $AAB$ systems.
The mass polarization term of the three-body kinetic-energy operator 
is important for evaluation of the $AA$ interaction strength\cite{H2002}.
For a bosonic-like system, when the  particle masses are related as $m_B \sim m_A$, the contribution of the mass polarization term 
to the three-body energy  can be evaluated as
$
\Delta =2E_{2}-E_{3}(V_{AA}=0),  \label{RK1}
$
where $\Delta>0$\cite{FG2002,H2002}. It means that $ |E_3(V_{AA}=0)|>|2E_2| $. For the $AAB$  systems having  isospin dependent $AB$ interaction we have obtained the opposite relation.

It has to be noted that, the $NN{\bar K}$ system can be  described by the ''bosonic-like'' model  in  which the isospin-dependence of $N{\bar K}$ interaction 
is removed by an averaging of the $N{\bar K}$ potential over isospin variables \cite{O,DHW08,DHW09,WG}. Resulted Faddeev equations describe a  three-body system where
the interaction between non-identical particles is given by isospinless potential presented  by a superposition of the singlet and triplet components of the $N{\bar K}$ potential. The averaged potential can be defined  by an algebraic transformation  \cite{FKSV17} for the Faddeev equations. We consider some properties of this averaged potential model and compare with the  isospin model.  

An alternative for the isospin  formalism based model  was proposed in the papers \cite{R16,DRS15,S17} as ''particle representation'' for the kaonic cluster $NN{\bar K}(S_{NN}=1)$. 
In this model, the $N\bar K$ subsystem is considered to be isospinless one to separate $pK^-$ and $n{\bar K}^0$ channels. The channel coupling was obtained by unitary transformation for the Schr\"{o}dinger equation described the $N{\bar K}$ pair.
Motivated by the model we develop the ''given charge formalism'' for $NN{\bar K}$ system to find a relation to  the 
 ''particle representation''.
\section{Formalism}
\label{sec:1}
\subsection{Faddeev equations for $AAB$ system}
\label{sec:2}
 The kaonic clusters $ppK^{-}$\ and $K^{-}K^{-}p$ represent  the three-body $AAB$ systems
with two identical particles. 
The total wave function of the $AAB$  system is decomposed into the sum of the Faddeev components $U$ and $W$ corresponding to the $(AA)B$ and $A(AB)$ types of rearrangements: $\Psi =U+W\pm PW$, where $P$ is the permutation operator for two identical particles.  In the expression for $\Psi$, the sign ''$+ $'' corresponds to two identical bosons, while the sign ''$- $'' corresponds to two identical fermions, respectively.
Each component is expressed by  corresponding Jacobi coordinates.
 For a three-body system with two identical particles the
set of the Faddeev equations is presented as a set of
two equations for the components $U$ and $W$ \cite{14}: 
\begin{equation}
\begin{array}{l}
{(H_{0}+V_{AA}-E)U=-V_{AA}(W\pm PW),} \\ 
{(H_{0}+V_{AB}-E)W=-V_{AB}(U \pm PW),}
\end{array}
\label{GrindEQ__1_}
\end{equation}%
where again the signs ''$+ $'' and ''$- $'' correspond to two identical bosons and fermions, respectively and $ H_{0}$
is the kinetic energy operator presented in the Jacobi coordinates for corresponding rearrangement. The wave function of the system $AAB$  is symmetrized with respect to
 two identical bosons, while it is antisymmetrized with respect to two identical fermions. 
In the presented work, we consider the $s$-wave approach for the $AAB$ systems. The 
total angular momentum $L=0$ and angular momenta in the subsystem $(AA)B$ and $A(AB)$ are equal zero.

\subsection{Isospin formalism for kaonic clusters}

In Eq. (\ref{GrindEQ__1_}), the Faddeev  components $U$ and $W$ of the total wave-function  are  expressed in terms of spin and isospin spaces.    
The  $NN{\bar K}$  system  is  a system with two identical particles  described by  Eq. (\ref{GrindEQ__1_}).
 In Eq. (\ref{GrindEQ__1_}), the Faddeev  component $U$ (and $W$) of the total wave-function  is  expressed in terms of spin and isospin spaces:
$$
U={\cal U} \chi_{spin}\eta_{isospin}.
$$

The  $NN{\bar K}$  system with the triplet isospin state of the pair of nucleons $I_{NN}=1$ is considered. 
The isospin basis for $NN{\bar K}$ system in the the state $I=1/2$ and $I^3=1/2$ can be written using the isospin functions: $\eta_{+-+}=\eta_+(1)\eta_-(2)\eta_+(3)$, $\eta_{-++}=\eta_-(1)\eta_+(2)\eta_+(3)$,  $\eta_{++-}=\eta_+(1)\eta_+(2)\eta_-(3)$. 
Where, for example,  $\eta_-(k)$ is eigenfunction of the isospin of $k$-th particle with projection of $-\frac{1}{2}$.
The three-body isospin basis for the configuration  $(1+2)+3$ includes two elements with diffident  isospins (single or triplet) of the (1+2) pair is written as
\begin{equation}
\label{e1} 
\begin{array}{l} 
\eta_1=\frac{1}{\sqrt{2}}( \eta_{+-+} - \eta_{-++}) , \;\;  {\rm singlet},\\
\eta_2=\sqrt{\frac{2}{3}}( \eta_{++-} -\frac{1}{2}\eta_{+-+}-\frac{1}{2}\eta_{-++}) , \;\; {\rm  triplet},\\
\end{array} 
\end{equation}
The basis for the configurations $(3+1)+2$ (and $(2+3)+1$) can be obtained from (\ref{e1}) by cyclical permutations of the isospin projections.
 
The spin states of the $NN{\bar K}$ system can be described by spin states of nucleon pair which can be  spin singlet or spin triplet.
The singlet  spin function $\chi^{s=0}({NN})$ is an asymmetrical function relatively the permutation of nucleons: $\chi^{s=0}({NN})~=~\frac{1}{\sqrt{2}}(\chi_{+-}- \chi_{-+})$, that provides the sign ''$+$'' in Eq. (\ref{GrindEQ__1_}).
The triplet spin function $\chi^{s=1}({NN})$ is symmetric, that gives the sign ''$-$'' in Eq. (\ref{GrindEQ__1_}).     

We employ the $s$-wave spin/isospin dependent $V_{AA}$ and $V_{AB}$ potentials. 
For $NN{\bar K}$ system, the separation of spin-isospin variables leads to the following form of the Faddeev equations: 
\begin{equation} 
\label{eq:1} 
\begin{array}{l} 
   (H_0+V_{AA}-E){\cal U}
  =-V_{AA}D(1+p){\cal W}, \\
   (H_0+V_{AB}-E){\cal W}=-V_{AB} 
  (D^T{\cal U}+Gp{\cal W}), 
\end{array} 
\end{equation} 
where  ${\cal W}$ is a column matrix with the singlet and triplet coordinate dependent parts of the Faddeev  component $W$, and the 
exchange operator $p$ acts on the particle coordinates. The component $U$ is presented by  single part ${\cal U} $ corresponding to isospin triplet state of $AA$ pair.  The  sign before operator $p$ in Eq. (\ref{GrindEQ__1_}) depends on spin state of the pair.    Within $s$-wave approach, the coordinate dependent part of $U$ corresponding to isospin singlet state is  dropped out from consideration due to the operator $(1-p)$,  which is appeared in right-hand side of  Eq. (\ref{GrindEQ__1_}). 

For ${\bar K}{\bar K}N$ and $NN{\bar K}$,  despite of the fact that there are two identical bosons and two identical fermions, respectively, due to  the symmetry of the spin-isospin configurations in  the
kaonic clusters, the $D$ and $G$ matrices in  (\ref{eq:1}) are the same and have the following form  \cite{K2015}:
\begin{equation}
\label{w2}
D =(-\frac{\sqrt{3}}{2},-\frac{1}{2}),  G=\left( 
\begin{array}{cc}
\frac{1}{2} & \frac{\sqrt{3}}{2} \\ 
\frac{\sqrt{3}}{2} & -\frac{1}{2}%
\end{array}%
\right), \quad
{\cal W}=\left( \begin{array}{rr}
{\cal W}^{s} \\ 
{\cal W}^{t} \\ 
\end{array} \right), \quad
{\cal U}= {\cal U}^{t}.
\end{equation}
The  superscripts $s$ and $t$ in  (\ref{w2})  denote the isospin singlet and isospin triplet coordinate dependent parts of the components $U$ and  $ W$.
For the ${\bar K}{\bar K}N$ kaonic cluster,  $V_{AA}=v^t_{{\bar K}{\bar K}}$ is the ${\bar K}{\bar K}$ potential in the triplet isospin state.
For the $NN{\bar K}$ cluster,  $V_{AA}=v^s_{NN}$ is the $NN$ potential in the singlet spin state. For both systems, one has to take $V_{AB}=diag\{v^s_{N{\bar K}},v^t_{N{\bar K}}\}$.

In the presented work, we used the $s$-wave Akaishi-Yamazaki
 (AY) \cite{YA07} and the simulating Hyodo-Weise (sHW)  effective potentials \cite{JK}  for $\bar{K} \bar{K}$ and  $N\bar{K}$ interactions,  
 which include the coupled-channel dynamics into a single
channel $N \bar{K}$ interaction.

The AY and sHW $N\bar{K}$ potentials are written in the form of one range Gaussian:
\begin{equation}
\label{AY_HW}
 V^{s(t)}_{N{\bar K}}(r)=V_0^{s(t)}exp((-r/b)^2),
\end{equation}
where for the AY potentials:
$V_0^{s}=-595.0$~MeV,  $V_0^{t}=-175.0$~MeV,  $b=0.66$~fm, and for the sHW potentials:
$V_0^{s}=-908.0$~MeV,  $V_0^{t}=-415.0$~MeV,  $b=0.47$~fm \cite{JK}.
The isospin triplet $\bar{K} \bar{K}$ potential is defined by the set: $V_0^{t}=104.0$~MeV,  $b=0.66$~fm  for the AY potential, and $V_0^{t}=313.0$~MeV,  $b=0.47$~fm \cite{KJ} for the sHW potential.
To describe the spin singlet nucleon-nucleon interaction  ($I=1$) we use the semi-realistic Malfliet-Tjon MT I-III \cite{MT} potential with the modification from Ref. \cite{MT1} :
$$
 V_{NN}(r)=(-513.968exp(-1.55r)+1438.72exp(-3.11r))/r ,
$$
where the strength parameters are given in MeV and  the range parameters are in fm$^{-1}$.
  
\subsection{Effect of isospin splitting of $AB$ potential}

Let us consider the $s$-wave approach  for the Faddeev equations (\ref{eq:1}) for the  $AAB$ system when 
 particles $A$ and $B$  interact via the isospin dependent   $V_{AB}$ potential, assuming that the interaction between two identical particles is omitted, 
therefore  $V_{AA}=0$. 

For the $NN{\bar K}$ and ${\bar K}{\bar K}N$ systems, Eq. (\ref{eq:1}) takes the form:
\begin{equation}
\label{eq:1comp1} 
\begin{array}{l} 
   (H_0+v^{s}_{N{\bar K}}-E){\cal W}^s
  =-v^{s}_{N{\bar K}}(\frac12 p{\cal W}^s+\frac{\sqrt{3}}2 p{\cal W}^t), \\
   (H_0+v^{t}_{N{\bar K}}-E){\cal W}^t=-v^{s}_{N{\bar K}}( \frac{\sqrt{3}}2 p{\cal W}^s- \frac12p{\cal W}^t) . 
\end{array} 
\end{equation}
Here, we have to take into account the significant difference between the isospin singlet and triplet components of the $N{\bar K}$ potential. The  isospin singlet component generates a deep bound state, while there is no bound state with the
triplet component. We can formulate that  $|v^{t}_{N{\bar K}}|<|v^{s}_{N{\bar K}}| $. 
The strengths of the  isospin components for the AY and sHW potentials are given in  (\ref{AY_HW}).

The Faddeev component  ${\cal W}^s$ of the total wave function of the kaonic clusters related to the isospin singlet interaction is dominant \cite{FKSV17}. 
We rewrite the equations (\ref{eq:1comp1})  to 
a simple form ignoring the small contribution coming from the isospin triplet component $W^t$:
$$
   (H_0+v^{s}_{N{\bar K}} -E){\cal W}^s=-v^{s}_{N{\bar K}}(\frac12p{\cal W}^s) 
$$
or 
\begin{equation}
\label{eq:1comp2a} 
\begin{array}{l} 
   (H_0+v^{s}_{N{\bar K}}+v^{s}_{N{\bar K}}p+ v^{s}_{N{\bar K}}(- \frac12p) -E){\cal W}^s=0. 
\end{array} 
\end{equation}
Here, taking into account the coefficient $\frac12$ we assume that the  term $\frac12v^{s}_{N{\bar K}}p$ is not a large perturbation.  
Ignoring this  term, the Eq. (\ref{eq:1comp2a})  is rewritten as
\begin{equation} 
\label{eq:10} 
\begin{array}{l} 
   (H_0+v^{s}_{N{\bar K}}+v^{s}_{N{\bar K}}p-E){\cal W}=0.
\end{array} 
\end{equation} 
It has been shown in  Ref. \cite{FKSV17}, that the last equation leads to the relation:
\begin{equation} 
\label{eq:10a} 
\begin{array}{l} 
  2E_2-\Delta-E_3(V_{AA}=0)=0,
\end{array} 
\end{equation} 
where $\Delta>0$ is evaluation for the mass polarization in three-body system \cite{H2002}. The 
mass polarization effect expressed by Eq. (\ref{eq:10a}) can be described by the mass polarization term $T_{mp}$, which is clearly seen  in the Schr\"{o}dinger equation   for the $AAB$ in the coordinate system of 
non-identical particles \cite{H2002}.   The  $\Delta$ is an approximation for the averaged value $<T_{mp}>$: $\Delta\approx <T_{mp}> $ when the contribution of the mass polarization term to the three-body energy $E_3(V_{AA}=0)$ is small: $\Delta/|E_3 (V_{AA}=0)|<1$. In the limit $m_B/m_A>>1$, the term  can be neglected and $2E_2=E_3(V_{AA}=0)$.

Taking into account Eqs. (\ref{eq:10}) and  (\ref{eq:10a}),  we obtain an approximation for Eq. (\ref{eq:1comp2a}):
\begin{equation}
\label{eq:1comp2b} 
\begin{array}{l} 
   2E_2-\Delta- \frac12<v^{s}_{N{\bar K}}p>-E_3(V_{AA}=0)=0. 
\end{array} 
\end{equation}
The matrix element $\frac12<v^{s}_{N{\bar K}}p>$ has negative value due to  attractive $v^{s}_{{\bar K}N}$
potential.
Thus, in Eq. (\ref{eq:1comp2b}), the terms $-\Delta$ and $- \frac12<v^{s}_{N{\bar K}}p>$ have opposite sings.
The interplay of the terms results in the two possible correlations
$|E_3(V_{AA}=0)|<2|E_2|$ or $|E_3(V_{AA}=0)|>2|E_2|$. Below, we show   that the relation   
\begin{equation}\label{eq:11}
|E_3(V_{AA}=0)|<2|E_2|
\end{equation}
takes place for the kaonic clusters.
It can be explained by strong attraction of the isospin singlet $N{\bar K}$ potential having a deep bound state. In this case, the Faddeev component $W^s$ is well factorized by the wave functions of the bound pairs $A_1B$ and $A_2B$ \cite{FKSV17}. Here, the indexes $1$ and $2$ distinguish the identical particles $AA$. 
At the same time, from the Eqs. (\ref{eq:10})-(\ref{eq:10a}) it is clear that the relation  $|E_3(V_{AA}=0)|>2|E_2|$ is satisfied for bosonic-like systems.

It has to be noted here that,  the relation (\ref{eq:11}) is not generally guaranteed (see Ref. \cite{FSV17}) 
for case  $m_B < m_A$, in the  systems complicated by spins or isospins. It should be again noted that the relation (\ref{eq:11})  has been  obtained under the condition $m_B/m_A>1$ and within the first order of the perturbation theory.  The condition of significant ''spin/isospin splitting'' of  the $AB$ potential is also necessary to be satisfied.
 
\subsection{Averaged potential model}
\label{subsec4}

In this section, we define the effective potential obtained by averaging of the
initial potential over the isospin variables. This averaging  produces the "isospinless" model for the  kaonic clusters. 
We will apply  the model for the $NN{\bar K}$ system when $S_{NN}=1$.

 The isospin averaged potential $V^{av}_{{\bar K}N}$ is defined as: 
\begin{equation}
V^{av}_{{\bar K}N}=\frac{3}{4}v_{{\bar K}N}^{s}+\frac{1}{4}%
v_{{\bar K}N}^{t}.  
\label{av_pot}
\end{equation}
Here, we use the isospin single and  triplet components $v_{{\bar K}N}^{s}$ and $v_{{\bar K}N}^{t}$ of the AY  $\bar{K} N$ potential.
This potential has a moderate attraction in comparison with the strong attraction in the $
{I=0}$ channel. The two-body threshold is changed to lower one and  is not related to the ${K^-}p$ bound state as $\Lambda $(1405). 

Using the isospin averaging, Eqs. (\ref{eq:1}) can be reduced to the scalar form by an algebraic transformation.
Taking into account that $W=D{\cal W}$, $V_{AB}^{av}=DV_{AB}D^T$ and $DD^T=1$, $DV_{AB}GD^T=V^{av}_{AB}$ one obtains 
\begin{equation} 
\label{eq:1av} 
\begin{array}{l} 
   (H_0^U+V_{AA}-E){\cal U}
  =-V_{AA}(1+p){\cal W} , \\
   (H_0^W+V^{av}_{AB}-E){\cal W}=-V^{av}_{AB} 
  (U+p{\cal W}). 
\end{array} 
\end{equation} 
In this case, one can evaluate  the mass polarization in the three-body system as
$
\Delta =2E^{av}_2-E^{av}_3(V_{AA}=0).
$
 Here, $E_2^{av}$ is the two-body energy for the $AB$ pair with the averaged potential and $E^{av}_3(V_{AA}=0) $
  is the three-body energy with the averaged potential when the ${AA}$ interaction is omitted. The value of $\Delta$ is positive one \cite{FKSV17}.
The averaged potential model was previously used in Refs. \cite{O,DHW08,DHW09,WG} for $NN{\bar  K}$ calculations  when $S_{NN}=1$ and $S_{NN}=0$. For the last case,
the averaged potential  is defined by Eq.  (\ref{av_pot}) where the superscripts $s$ and $t$ are exchanged.

\subsection{Isospin given charge formalism}

The systems $NN{\bar K}(I_{NN}=1)$ and  $NN{\bar K}$($s_{NN}=0$) is separate when the $N{\bar K}$ interaction does not include an isospin mixing component. The isospin bases are  orthogonal due to chose the total isospin projections which is motivated by the ''isospin charge'' set of particles: $ppK^-$ and $pnK^-$: $(++-)$ and $(+--)$.
The isospin functions $\eta_{+-+}$, $\eta_{-++} $ and $\eta_{++-}$ represent  new isospin basis $\tau$ with the elements $\tau_1$, $\tau_2$, $\tau_3$, respectively.  The $\tau$-basis elements  not relate to the fixed isospin of pair. 
We will describe these basis as ''given charge basis''. 
To obtain matrix of transformation between 
both bases we have to add an additional element to $\eta$-basis (\ref{e1}). The element relates to the isospin state of the  $NN{\bar K}$ system with total isospin equal to $3/2$ and projection $1/2$ (or $-1/2$). 
For the configuration $(1+2)+3$, this basis element is written as 
\begin{equation}
\label{e2} 
\begin{array}{l} 
\eta_3=\frac{1}{\sqrt{3}}( \eta_{+-+} +\eta_{-++} +\eta_{++-}) , \;\;  {\rm triplet}.\\
\end{array} 
\end{equation}
The pair potentials $NN$ and ${\bar K} N$ have diagonal  representation in the basis (\ref{e1}),  (\ref{e2}):
\begin{equation}
\label{V}
V=diag\{v^s,v^t,v^t\}.
\end{equation}

The matrix of transformation of the $\eta$ and $\tau$ bases is given by flowing relation:
\begin{equation}
\eta=S{\tau},
\end{equation}
where 
\begin{equation}
\label{beta}
\tau=({\tau}_1, {\tau}_2, {\tau}_3)^T,   
 \qquad {\tau}_1= \eta_{+-+}, \qquad
{\tau}_2 = \eta_{-++}, \qquad
{\tau}_3 = \eta_{++-},
\end{equation}
and
\begin{equation}
\label{S}
S = \left(\begin{array}{ccc}
 \frac{1}{\sqrt{2}} & -\frac{1}{\sqrt{2}}&0\\
-\frac{1}{\sqrt{6}} & -\frac{1}{\sqrt{6}}&\sqrt{\frac{2}{3}} \\
\frac{1}{\sqrt{3}} & \frac{1}{\sqrt{3}} & \frac{1}{\sqrt{3}}
 \end{array}\right).
\end{equation}
The matrix $S$ is unitary: $S^TS=I$. 

In the ''given charge''  basis (\ref{beta}), the matrix representation for potentials has non-diagonal elements: 
\begin{equation}
\label{SVS}
S^TVS = \left(\begin{array}{ccc}
\frac12(v^t+v^s) &\frac12(v^t-v^s)&0\\
\frac12(v^t-v^s)&\frac12(v^t+v^s)&0 \\
 0& 0 & v^t
 \end{array}\right)=\left(\begin{array}{ccc}
V^+&V^-&0\\
V^-&V^+&0\\
 0& 0 & v^t
 \end{array}\right),
\end{equation}
where $V^+=\frac12(v^t+v^s) $ and $V^-=\frac12(v^t-v^s) $.

Let us to define the cyclical permutation operators $P_c$. The $\eta$-bases  related to the configuration $(3+1)+2$ and $(2+3)+1$ are $\tilde{\eta}=P_c\eta$ and $\tilde{\tilde{\eta}}=P_cP_c\eta$. 
Taking into account the Eq. (\ref{GrindEQ__1_}), we  write the matrix representation of operators $I$ (or $P_c$) and $P$ in the bases $\eta$ and $\tilde{\eta}$ as following 
\begin{equation}
<\eta|I| \tilde{\eta}>=I^{(1,2)}=\left(
\begin{array}{ccc}
 -\frac12 & -\frac{\sqrt3}2 & 0 \\
 \frac{\sqrt3}2 & -\frac12 & 0 \\
 0 & 0 & 1
\end{array}\right), 
<\eta|P| \tilde{\eta}>=P^{(1,2)}=\left(
\begin{array}{ccc}
\frac12 & \frac{\sqrt3}2 & 0 \\
\frac{ \sqrt3}2 & -\frac12  & 0 \\
 0 & 0 & 1
\end{array}\right),
\end{equation}
  \begin{equation}
<\tilde{\eta}|I| \eta>=I^{(2,1)}=\left(
\begin{array}{ccc}
 -\frac12  &  \frac{\sqrt3}2 & 0 \\
 -\frac{\sqrt3}2 & -\frac12  & 0 \\
 0 & 0 & 1
\end{array}\right),  
<\tilde{\eta}|P| \tilde{\eta}>=P^{(2,2)}=\left(
\begin{array}{ccc}
 \frac12  & -\frac{\sqrt3}2 & 0 \\
 -\frac{\sqrt3}2 & -\frac12  & 0 \\
 0 & 0 & 1
\end{array}\right).
\end{equation}
The unitary transformation given by $S$ matrix of Eq. (\ref{S}) leads to the matrices:
\begin{equation}
\label{T1}
S^TVI^{(1,2)}S=\left(
\begin{array}{ccc}
 0 &V^+ & V^-\\
0 & V^- & V^+ \\
v^t & 0 & 0
\end{array}\right), \quad 
S^TVP^{(1,2)}S=\left(
\begin{array}{ccc}
 0 &V^- & V^+\\
0 & V^+ & V^- \\
v^t & 0 & 0
\end{array}\right),
\end{equation}
\begin{equation}
\label{T2}
S^TVI^{(2,1)}S=\left(
\begin{array}{ccc}
 V^-& 0 &V^+ \\
V^+ & 0 &V^- \\
 0 &  v^t & 0
\end{array}\right), \quad 
S^TVP^{(2,2)}S=\left(
\begin{array}{ccc}
 V^+& 0 &V^- \\
V^- & 0 &V^+\\
 0 &  v^t &0
\end{array}\right).
\end{equation}
 Thus,    the unitary transformation $S$ results in new set of the Faddeev equations instead  Eq.~(\ref{eq:1}). New set  
  includes an additional equation with isospin triplet potential and relates to expansion of the $\eta$-basis by  the  isospin channel $I=3/2$.
   
 Similar transformation have been proposed in Ref. \cite{R13} within ''particle representation''  for  $NN{\bar K}$($s_{NN}=1$) system. Within the model, the author of \cite{R13} represented elements of the given charge basis (\ref{beta})
 as the physical channels ${\bar K}^0nn$, $K^-pn$, $K^-np$ taking into account possible particle transition. Within such interpretation, the  non-diagonal elements $V^-$ of the matrix representation (\ref{SVS}) of the $N{\bar K}$ potentials  in $\tau$-basis were considered as a ''channel coupling''.
 Obviously, the ''channel coupling'' appeared after unitary transformation of the two-body equation for $N{\bar K}$ is not related with a new physical effect.
  
 The similar channel interpretation   one can found in Ref. \cite{O17,H17}.  The coupled channel Schr\"{o}dinger  equation was written as 
 $$
 (H_0-E+\left(
\begin{array}{cc}
 V_+ &V_- \\
V_- &V_+
\end{array}\right) )\phi=0,
 $$
where  $\phi=(\phi_1,\phi_2)^T$ and $\phi_1$ ($\phi_2$) corresponds to $K^-p$ (${\bar K}^0n$) state of $N{\bar K}$,  $V_+=V^++V^-$,  $V_-=V^+-V^-$.
The unitary transformation $t=\left(\begin{array}{cc}
 -\frac{1}{\sqrt{2}} & \frac{1}{\sqrt{2}}\\
   \frac{1}{\sqrt{2}} & \frac{1}{\sqrt{2}}
 \end{array}\right)$ separates the channels as following 
$$
 (H_0-E+\left(
\begin{array}{cc}
 V^+ & 0 \\
0&V^-
\end{array}\right) ){\tilde \phi}=0.
 $$
 The last equations mean a redefinition for singlet and triplet components of $N{\bar K}$ potential. The potentials $v^s$ and $v^t$ did not clarify  in this model due to the 
 components $\phi_1$ and $\phi_2$ are represented as $(-+)$ and $(-+)$ and the components ${\tilde \phi}_1$ and ${\tilde \phi}_1$ as symmetric and antisymmetric combinations of $\phi_1$ and $\phi_2$.
  
  Let us to consider the $V_{AA}$ and $V_{AB}$  potentials without the isospin dependence. We can assume that $v^s=v^t=v$.
 The  non-diagonal elements in Eqs.   (\ref{SVS}), (\ref{T1}),  (\ref{T2}) will be equal to zero.  The corresponding set of the Faddeev equations can be reduced using  the matrix transformation: $ \left(\begin{array}{ccc}
 0 & 1&0\\
0& 0&-1\\
1 & 0& 0
 \end{array}\right)$, and we obtain the ''isopin-less'' model, like the model given by Eq. (\ref{eq:1av}). 
  Such approach was employed  in Ref. \cite{FSV16} to describe   the $nnp$ and $nnp$ systems  as an ''isospinless'' systems. The final equations were obtained  by using a spin basis and taking into account the spin-splitting nucleon-nucleon potential.
  
One can make one more unitary transformation $T$ for the Eq. (\ref{T1} ) and (\ref{T2}). The $T$ is defined by the matrix
  \begin{equation}
T = \left(\begin{array}{ccc}
 -\frac{1}{\sqrt{2}} & \frac{1}{\sqrt{2}}&0\\
   \frac{1}{\sqrt{2}} & \frac{1}{\sqrt{2}}&0\\
0 & 0& 1
 \end{array}\right).
\end{equation}
 The new matrix representation of the potentials has the same form (\ref{V}) as it had for the $\eta$-basis. The operators $I$ and $P$ are presented as 
  \begin{equation}
  \label{1}
I^{(1,2)}=\left(
\begin{array}{ccc}
 -\frac12 & -\frac12 & \frac1{\sqrt2} \\
  \frac12 & \frac12 & \frac1{\sqrt2} \\
-\frac12 & \frac12 & 0 
\end{array}\right), \qquad
P^{(1,2)}=\left(
\begin{array}{ccc}
\frac12 & \frac12 & -\frac1{\sqrt2} \\
  \frac12 & \frac12 & \frac1{\sqrt2} \\
-\frac12 & \frac12 & 0 
\end{array}\right),
\end{equation}
  \begin{equation}
  \label{2}
I^{(2,1)}=\left(
\begin{array}{ccc}
-\frac12 & \frac12 & -\frac1{\sqrt2} \\
 -\frac12 & \frac12 & \frac1{\sqrt2} \\
\frac12 & \frac12 & 0
\end{array}\right),  \qquad
P^{(2,2)}=\left(
\begin{array}{ccc}
\frac12 & -\frac12 & \frac1{\sqrt2} \\
 -\frac12 & \frac12 & \frac1{\sqrt2} \\
\frac12 & \frac12 & 0
\end{array}\right).
\end{equation}
In this case, the right-hand side of the Faddeev equations  mixes the components  related to the  singlet and triplet potentials.  The type of symmetry of  spin wave function of nucleon pair defines the set of the Faddeev equations. For singlet spin state with antisymmetric spin function, the equation with isospin singlet $NN$ potential is dropped out due to the factor $(1-p)$ in the right side of the Faddeev equations for ${\cal U}$ component. The symmetric spin function corresponding to $s_{NN}=1$ drops out the second equation for  ${\cal U}$ component, which is associated to the isospin triplet  $NN$ potential.  In this case, the first equation  includes  the isospin singlet/spin triplet $NN$ state, realized as the deuteron.

The quantum numbers of  $NN{\bar K}$ system are determined as $s_{NN}=0$, $I=1/2$, $I_{z}=1/2$ for $pp{K^-}$ and as  $s_{NN}=1$, $I=1/2$, $I_{z}=-1/2$ for $np{K^-}$. The states can be described separately because the  $\tau$-bases  are orthogonal: $(\{+-+\},\{-+-\})=0$, $\dots$. Obviously, the condition  that $N{\bar K}$ potential does not mix the  corresponding bases is necessary. With this condition,    $<ppK^-|V_{AB}|npK^->=0$  due to the orthogonality. In opposite to the isospin model, the model \cite{DHW08} with the  averaged   $N{\bar K}$ potential yields  non-zero matrix element. This coupling effect takes  place  in the averaged potential model as well as in an  ''isospinless'' model. 

One more difference of the model \cite{DHW08} and the isospin model is appropriate to noted here.
One can consider the $NN{\bar K}$ system in these two states  as  $ppK^-$/$npK^-$  two-level system as it was proposed in Ref. \cite{DHW08}. The  level anti-crossing formalism requests that two levels have to be bound states. However, the system  $npK^-$ is unbound within the isospin formalism and is bound in the model developed in Ref. \cite{DHW08}. The similar calculations \cite{UHO11} used  continuous spectrum function in the  $npK^-$ channel within the ''sharp resonance'' approximation. The contribution of this channel has been evaluated as negligible.

\section{Numerical  Results}

We have calculated the ground state energy for ${\bar K}{\bar K}N$  and  $NN{\bar K}$ systems for a complete set of the potentials and under the condition $V_{AA}=0$. The Faddeev equations (\ref{eq:1}) were numerically solved using  the cluster reduction method \cite{CRM}.
The  results are presented in Table \ref{t1a}. For the both potentials AY and sHW,  the relation $2E_2-E_3(V_{AA}=0)<0$ is satisfied.
The three-body binding  energy $|E_3|$ is smaller than the value $|E_3(V_{AA}=0)|$ for the ${\bar K}{\bar K}N$ system and is larger for  $NN{\bar K}$,
due to the repulsive and weak attractive $V_{AA}$ potential, respectively. For the $NN \bar K$ system $|E_3|$  increases  from the $|E_{3}(V_{AA}=0)|$ value for 3~MeV,
when the AY potential is used, that is about 24\% in the binding energy $|E_3|$ measured relatively to the two-body threshold.
Obtained results for the ${\bar K}{\bar K}N$  and  $NN{\bar K}$ systems are comparable with the results of calculations performed within different approaches \cite{FKSV17}.   For  example, calculated values $|E_3|$ reported in Ref. \cite{RS14}  are  47--54~MeV for the phenomenological  ${\bar K}N$ potentials.
Including the possible physical channels into consideration may result in a small increase of the binding energy  $|E_3|$  relatively to $|E_{3}(V_{AA}=0)|$ (see for example \cite{D17}). However, such approaches cannot reach the values above 60~MeV. 

\begin{table}[t]
\caption{
Ground state energies $E_3$ of the ${\bar K}{\bar K}N$  and $NN{\bar K}$  systems with the AY and sHW  potentials for the
${\bar K}N$ and ${\bar K}{\bar K}$  interactions and the MT I-III potential for the $NN$ interaction. 
The difference $\delta$ of  the two-body  $2E_2$ and three-body  $E_3$ energies, $\delta=2E_{2}-E_{3}$, is presented. 
The energies are given in MeV.
}
\label{t1a} 
{\begin{tabular}{llccc} \hline\noalign{\smallskip}
System&   Potentials    &$E_{2}$  &  $E_{3}$& $\delta$  \\  \noalign{\smallskip}\hline\noalign{\smallskip}
 $K^-K^-p$ & $V_{{\bar K}{\bar K}}=0$, AY& -30.30&-35.2  &-25.3\\   
                 & AY, AY & &-31.7 & -28.9 \\ 
                & sHW, sHW  &-11.16 & unbound & --    \\
                &$V_{\bar K \bar K}=0$, sHW   & &-12.2  &  -10.1\\                                
                 \hline
$ppK^-$   & $V_{NN}=0$, AY & -30.30&-42.9  & -17.6 \\                          
                &  MT I-III, AY& &-46.0 & -14.6 \\  
                & $V_{NN}=0$, sHW&-11.16  & -17.1  & -5.20\\                          
                &  MT I-III, sHW & &-21.0  & -1.3\\           \noalign{\smallskip}\hline   
\end{tabular} }
\end{table} 

\begin{figure}
\resizebox{0.55\textwidth}{!}{%
\includegraphics{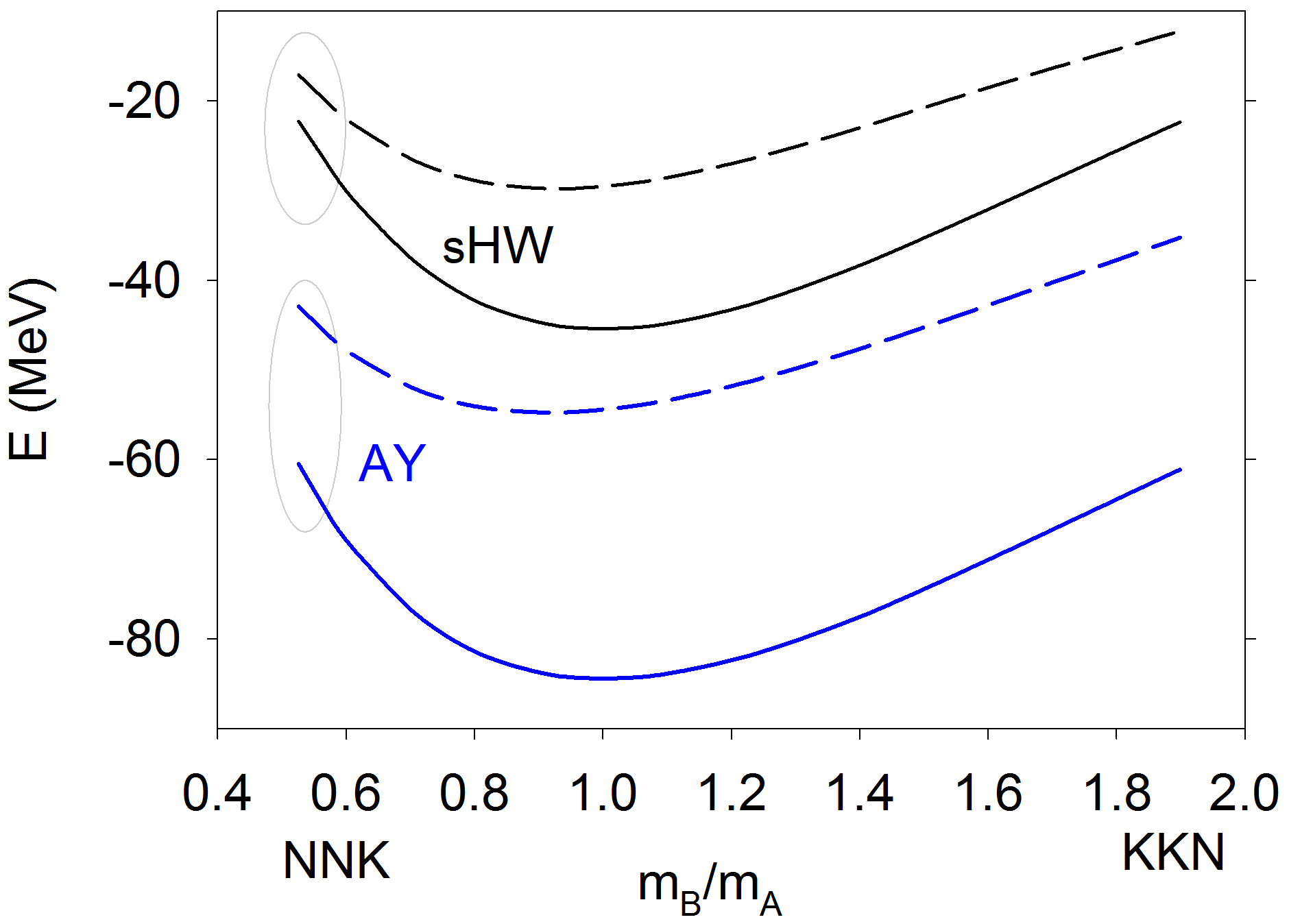}
}
\caption{  
The evolution of $NN \bar K$  to ${\bar K \bar K} N$ trough the mass transformation (\ref{eq:par})  $m_N \to m_{\bar K}$ and $m_{\bar K} \to m_N$ when $m_N+m_{\bar K}=const$. The energies  $2E_2$ (solid line) and $E_3(V_{AA}=0)$ (dashed line) are shown as functions of the ratio $m_B/m_A$ for AY and sHW $\bar K N$ potentials.  The $NN \bar K$ (${\bar K \bar K}N$) system corresponds to the value of 0.526 (1.90) for the ratio $m_B/m_A$. 
}
\label{fig1}
\end{figure}

To illustrate the relation $2E_2-E_3(V_{AA}=0)<0$, in Fig. \ref{fig1}, we show the evolution of the $NN \bar K$  to ${\bar K \bar K} N$ trough the mass transformation  $m_N \to m_{\bar K}$ and $m_{\bar K} \to m_N$ when $m_N+m_{\bar K}=const$.   A parametric representation for the mass change is given by the formula:
\begin{equation}
\label{eq:par} 
   m^\xi_N= (1- \frac{m_{\bar K}}{m_N}\xi)m_N, \quad m^\xi_{\bar K}= (1+\xi)m_{\bar K}, 
\end{equation}
where  $0< \xi < m_N/m_{\bar K}$.
 The ratio $m_B/m_A$ has the value of  0.526 for 
$NN \bar K$  system and  the value of 1.90 for  ${\bar K \bar K}N$ system.
The relation (\ref{eq:11}) is well satisfied for AY potential when $m_B/m_{A}<1$. For the sHW potential which is weaker, the relation is still  satisfied. It is clear that subsequently weaker ${\bar K}N$ potential could violate the relation (\ref{eq:11}). Thus, the existence of deep bound state of nucleon and kaon is necessary 
for the relation (\ref{eq:11}). For $AAB$ systems with weak spin/isospin dependent $AB$ potential  the relation (\ref{eq:11}) is not guaranteed when $m_B/m_{A}<1$.
This conclusion is also supported by the calculations for the $NN\Xi$ and $\Xi\Xi N$ systems presented in Ref. \cite{FSV17}.
The relation (\ref{eq:11}) will be satisfied for any ${\bar K}N$ potential when $m_B /m_A>1$, since the contribution of the mass polarization energy decreases to zero when $m_B /m_A>>1$. 

\begin{figure}
\resizebox{1.0\textwidth}{!}{%
\includegraphics{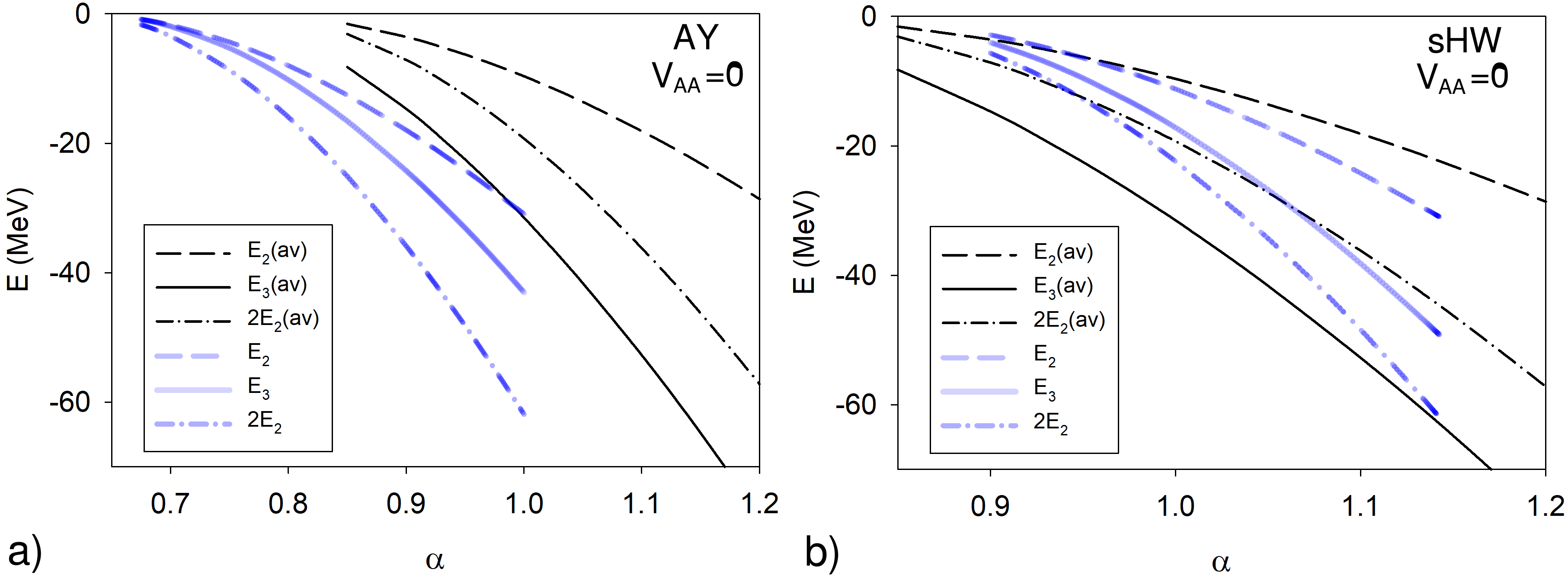}
}
\caption{  $NN \bar K$ system:
the energies $E_2$ (dashed line),  $2E_2$ (dot-dashed line) and $E_3$ (solid line) are shown as functions of the scaling factor  $\alpha$:  a) for AY and averaged (av) AY $\bar K N$ potentials, b) for sHW and averaged (av) AY $\bar K N$  potentials. The $AA$ potential acting between identical particles is neglected, $V_{AA}=0$.
}
\label{fig2}
\end{figure}

\begin{figure}
\resizebox{1.0\textwidth}{!}{%
\includegraphics{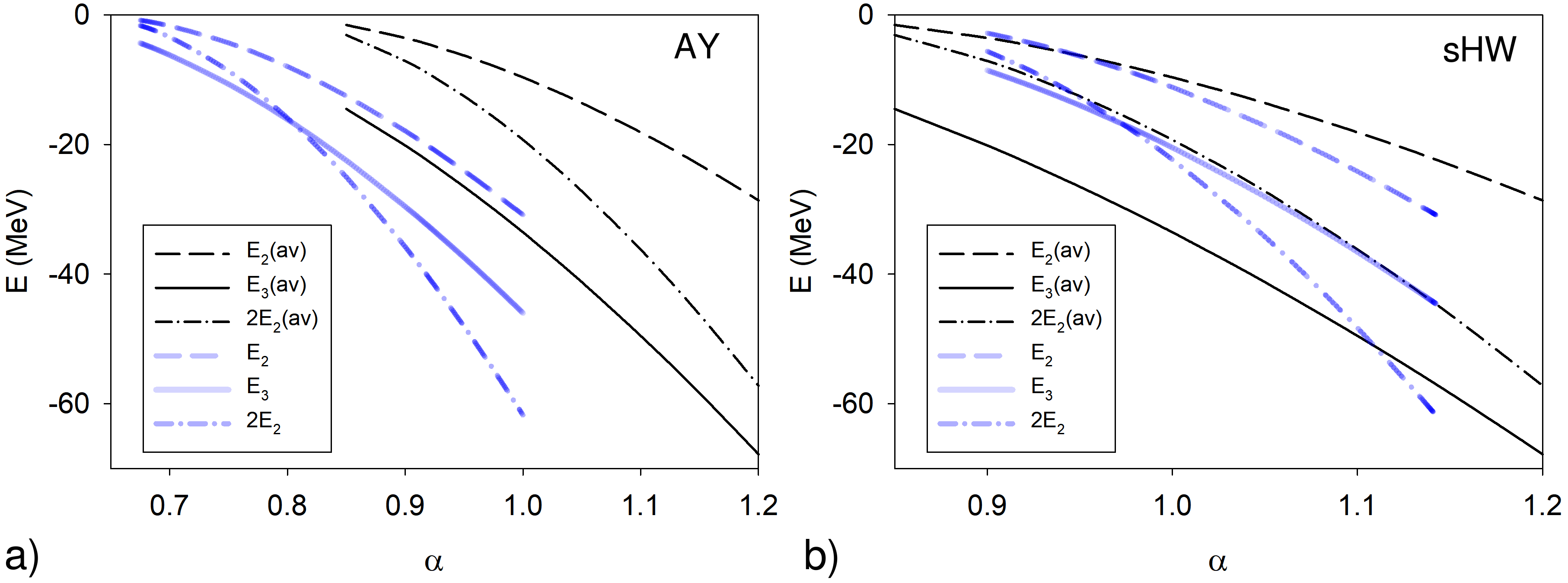}
}
\caption{  $NN \bar K$ system:
the energies $E_2$ (dashed line),  $2E_2$ (dot-dashed line) and $E_3$ (solid line) are shown as functions of the scaling factor  $\alpha$:  a) for AY and averaged (av) AY $\bar K N$ potentials, b) for sHW and averaged (av) AY $\bar K N$  potentials. 
}
\label{fig3}
\end{figure}

We illustrate the existence the lower bounds for the ground state  energy of the  $NN \bar K$  system in Fig.  \ref{fig2} and \ref{fig3} using the AY, sHW and averaged (av) potentials for $\bar K N$ interaction.   The energies $E_2$,  $2E_2$  and $E_3$  are shown as functions of the scaling factor  $\alpha$ 
which controls 
the strength of interaction between non-identical particles: $V_{AB}\to \alpha V_{AB}$. 
The case, when the $AA$ potential acting between identical particles is neglected, $V_{AA}=0$, is presented in Fig.  \ref{fig2}. One can see that 
the relation (\ref{eq:11}) is well satisfied for both models with the AY and sHW potentials. 
The isospin-less model with averaged (av) potential demonstrates opposite relation. Here, the mass polarization  depends weakly on strength of the  $AB$ potential and  $2E_2-E_3(V_{AA}=0)\approx const$.

The situation is slightly altered when the $AA$ interaction is included to the calculations as it is shown in Fig. \ref{fig3}.
The attractive  $NN$ interaction affects the $E_3$ curve which becomes  lower than is one in  Fig. \ref{fig2}. The relation  (\ref{eq:11}) is well satisfied for the large values of the two-body ground state energy, $E_2> 10$~MeV. In the sector of weak $AB$ potential, the competition between the terms $-\Delta$ and $- \frac12<v^{s}_{N{\bar K}}p>$ of Eq. (\ref{eq:1comp2b})  leads to domination of the first one due to adding the weak $NN$ attraction to the $-\Delta$ and the opposite relation $|E_3(V_{AA}=0)|>2|E_2|$ is satisfied.

Note here that for the model with the averaged (av) potential, the $E_3$ becomes to closer to $2E_2$ in the sector of  large strength  of $AB$ potential. It can be explained by the core effect of the $NN$ potential which is only appeared for the isospin-less model. The repulsion of the core plays a role when 
three-body system is very compact. Strong  $N\bar K$ interaction provides the repulsive effect of $NN$ potential.
 \begin{table}[hb]
\caption{
Ground state energy $E_3$ of the  $NN{\bar K}$  system  calculated within the averaged potential and  isospin models.
The energies are given in MeV.
}
\label{t2} 
{\begin{tabular}{cccccc} \hline\noalign{\smallskip}
               Model          & \cite{WG}  & \cite{DHW09}& Our & \cite{YA07}& \cite{K2015}    \\  
                                 & KWW  &  AY+T & AY+MT & AY+T & AY+AV14 \\
               \noalign{\smallskip}\hline\noalign{\smallskip}
 Averaged potential   &-35.5 & -39.1 &-33.6&  & \\                          
Isospin  &   &  & -46.0& -48 & -47.34 \\  
                       \noalign{\smallskip}\hline   
\end{tabular} }
\end{table} 

 The deference between calculations for  the $NN \bar K$  binding energy within isospin and averaged potential models is shown in Table. \ref{t2}.
 We compare the results of the different authors from Refs.  \cite{WG,DHW09} for the averaged AY potential model  and ones obtained in the isospin model  \cite{YA07,K2015}. Our results correspond to the calculations  with the AY and MT-I-III  potentials for both models.
 In Ref.  \cite{WG}, the   KWW potential set was used. The AY and  Tamagaki potentials were applied in Refs.  \cite{DHW09} and \cite{YA07}. The AY and  AV14 potentials were used in  \cite{K2015}.
 The energy $E_3$ of the averaged potential model  is always larger  relatively one calculated in the isospin model. This could  be expected due to  more higher position of two-body threshold $E_2$ of the  averaged potential model.

\section{Conclusions}
 The  kaonic clusters ${\bar K}{\bar K}N$ and $NN{\bar K}$  are examples of three-body $AAB$ system  with an isospin dependent $AB$ interaction.
The relation between three, $E_3$, and two, $E_2$, - body ground state energies was studied for the kaonic clusters.
It was shown that the  "isospin splitting" of the $AB$ interaction  leads to the  relation $ |E_3(V_{AA}=0)|<|2E_2|$. 

Based on the Faddeev calculations 
for the  kaonic clusters, we have found that  this relation is satisfied for the case of the AY $N{\bar K}$  potentials having one range Gaussian form when mass ration is  $0.5 <m_B/m_A<2$.
Thus, we have obtained the lower bound for $E_3(V_{AA}=0)$ which can be reached by using this phenomenological  isospin-dependent potential.
For  $NN{\bar K}$ cluster, we evaluated  $|E_3(V_{NN}=0)|\approx$43~MeV.
  $|E_3|$ has to be
 larger than  $|E_3(V_{NN}=0)|$, due to the weak attraction of the  $NN$ force. However, 
 calculated value of $|E_3|$  is smaller than $|2E_2|$ ($\sim $60~MeV) and is significantly smaller  the ''experimentally motivated value'' of 100~MeV.  
 
 The model based on the isospin averaged  $N{\bar K}$ potential was considered. This ''isospinless'' model demonstrates the opposite relation:  $\left\vert E_{3}(V_{AA}=0)\right\vert~>~2\left\vert E_{2}\right\vert$. The averaged potential changes two-body threshold $|E_2|$ to a smaller value. Thus, three-body binding energy  $|E_3|$  is significantly different  comparing to one calculated within the   isospin model.   
 The coupling  between the states  $s_{NN}=0$ and $s_{NN}=1$ of  the $NN{\bar K}$ system occurred in the averaged potential model \cite{DHW08}  looks  as artificial. Generally, one can conclude, that the averaged potential model is a rough approximation for the isospin model.
 
The isospin ''given charge formalism'' for  $NN{\bar K}$ cluster was proposed. This formalism was motivated by the ''particle representation'' which has been developed in a number of papers.   We shown that  the ''channel interpretation'' of the particle model is not appropriate to describe a possible particle transition. The ''channel  coupling'' is appeared as a result of  unitary transformation of the $\eta$ - isospin basis. 
 
\section*{Acknowledgments}
We thank Prof. R.Ya. Kezerashvili, Prof. V.M. Suslov, Prof. A. Gal and Prof. M.A. Braun for useful discussion.
This work is supported by 
the National Science Foundation (HRD-1345219). 
\end{document}